# Ab Initio Exchange Interactions and Magnetic Properties of Intermetallic Compound $Gd_2Fe_{17-x}Ga_x$.

E.E.Kokorina[1, a], M.V.Medvedev[1,b] and I.A.Nekrasov [1,c]

[1]Institute of Electrophysics, Ural Branch of Academy of Sciences,

Ekaterinburg, Amundsena,106, Russia

[a] kokorina@iep.uran.ru, [b] medvedev@iep.uran.ru, [c]nekrasov@iep.uran.ru

**Keywords:** intermetallic, Curie temperature, exchange parameters

**Abstract.**
Intermetallic compounds $R_2Fe_{17}$ are perspective for applications as permanent magnets. Technologically these systems must have Curie temperature $T_c$ much higher than room temperature and preferably have easy axis anisotropy. At the moment highest $T_c$ among stoichiometric $R_2Fe_{17}$ materials is 476 K, which is not high enough. There are two possibilities to increase $T_c$: substitution of Fe ions with non-magnetic elements or introduction of light elements into interstitial positions. In this work we have focused our attention on substitution scenario of Curie temperature rising observed experimentally in $Gd_2Fe_{17}$-$xGa_x$ (x=0,3,6) compounds. In the framework of the LSDA approach electronic structure and magnetic properties of the compounds were calculated. Ab initio exchange interaction parameters within the Fe sublattice for all nearest Fe ions were obtained. Employing the theoretical values of exchange parameters Curie temperatures $T_c$ of $Gd_2Fe_{17}$-$xGa_x$ within mean-field theory were estimated. Obtained values of $T_c$ agree well with experiment. Also LSDA computed values of total magnetic moment coincide with experimental ones.

## Introduction

For using the intermetallic compound $R_2Fe_{17}$ as permanent magnet it should have higher Curie temperature $T_c$ and easy axis anisotropy. There are several ways to achieve those goals [1]. First – to substitute Fe by nonmagnetic ions, second – to add interstitial C or N atoms. Problem of $T_c$ increase was investigated on the example of Ga substitution into rhombohedral compound $Gd_2Fe_{17}$.

## Crystal structures and local magnetic moments

The intermetallic compound $Gd_2Fe_{17}$ can crystallize both in the rhomohedral $Th_2Zn_{17}$ and the hexagonal $Th_2Ni_{17}$ crystal structure. According to experimental results [2,3] Curie temperatures $T_c^{rh}$ is higher then $T_c^{hex}$, so the rhombohedral phase is more interesting. In all our further investigations we focus our attention on the rhombohedral phase (see Fig.1) (space group R3m – no.166 in International Tables of Crystallography).

Main structural blocks are irons hexagons layers. The Fe(18f) hexagons (let us denote Fe(18f) as Fe3) contain the Gd ions in positions 6c either in center, or alternate with empty hexagons. Intermediate layers without Gd contain ions Fe(9d)=Fe2. Interlayer Fe(6c)=Fe1 ions at the dumbbell position lower the symmetry of nearest hexagon down to Fe(18h)=Fe4.

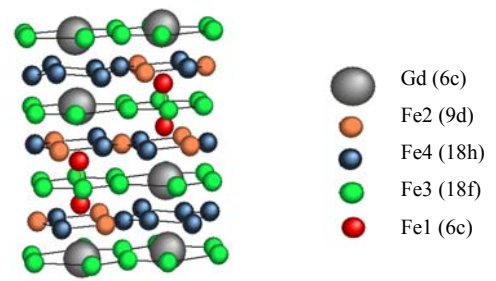

Fig1. Rhombohedral structure of $Gd_2Fe_{17}$.

Let us introduce Ga in the lattice. With Ga atoms substitution the unit cell volume increases. Lattice parameters for $Gd_2Fe_{17}$-$xGa_x$ [4] are shown in Table 1.

In which positions will Ga substitute the Fe atoms (6c(Fe1), 9d(Fe2), 18f(Fe3) or 18h(Fe4)) depends on concentration of Ga. For x<2 substitution starts from Fe4 and at x~3 the Fe3 positions added, and at x~6 the amounts of positions Fe4 and Fe3 become equal [5,6]. In our further calculations we restrict ourselves to the substitutions into Fe3 positions as the most symmetric ones (see Fig. 2,3). Other configurations of Ga substitutions are to be published.

Table1. Lattice parameters and $T_c$ values for $Gd_2Fe_{1-x}Ga_x$ with x=0,3,6.

|  | $Gd_2Fe_{17}$ | $Gd_2Fe_{14}Ga_3$ | $Gd_2Fe_{17}Ga_6$ |
|---|---|---|---|
| a [A] | 8.538 | 8.616 | 8.697 |
| c [A] | 12.431 | 12.559 | 12.648 |
| $T_c$(exp.[4]), [K] | 465 | 610 | 410 |
| $T_c$(calc.),[K] | 429 | 521 | 381 |

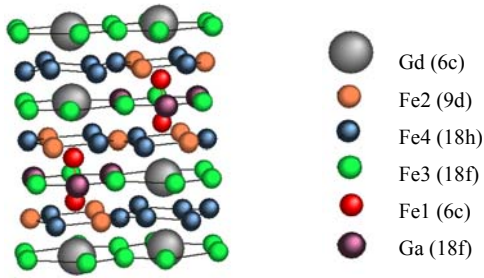

Fig.2. $Gd_2Fe_{14}Ga_3$, with Ga in positions of Fe3(18f).

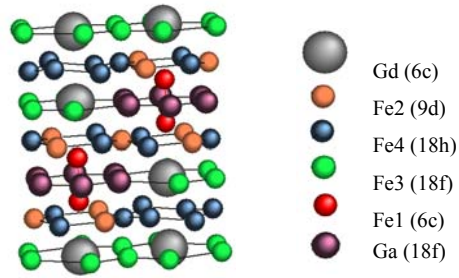

Fig.3. $Gd_2Fe_{17}Ga_6$, with Ga in positions of Fe3(18f).

The electronic structure of rhombohedral $Gd_2Fe_{17-x}Ga_x$ was calculated within the LSDA+U method [7] using package TB-LMTO-ASA (tight-binding, linear muffin-tin orbitals, atomic sphere approximation) [8]. Atomic spheres radii were chosen as R(Gd)=2.86 a.u. and R(Fe)=2.62 a.u. The division of Brillouin zone is 6*6*6.

In the LSDA+U method Coulomb interaction was taken into account via parameters of direct U and exchange J Coulomb interactions of 4f-electrons in Gd (since those correlation effects are much stronger than for the Fe 3d shell). For $Gd_2Fe_{17}$ constrained LDA method [9] gives $U_{Gd}$=6.7 eV and $J_{Gd}$=0.7 eV that is in agreement with the values of Gd metal [7]. In the LSDA+U calculations for all concentrations of Ga these values were used for Gd ions. We believe that greater part (not all) of electronic interaction effects is taken into consideration for Fe 3d shell in the frame of LSDA approximation. It is well enough justified since responsible Stoner parameter for Fe 3d shell $J_{Fe}^S$~1 eV is close to $U_{Fe}$.

Local magnetic moments for various Fe and Gd sites obtained in the LSDA+U calculations are listed in Table 2.

Table2. Calculated values of local and total magnetic moments for $Gd_2Fe_{1-x}Ga_x$ with x=0,3,6.

| Site | M[$\mu_B$] for $Gd_2Fe_{17}$ | M[$\mu_B$] for $Gd_2Fe_{14}Ga_3$ | M[$\mu_B$] for $Gd_2Fe_{17}Ga_6$ |
|---|---|---|---|
| Gd (6c) | -7.17 | -7.21 | -7.31 |
| Fe(6c)=Fe1 | 2.19 | 2.07 | 2.04 |
| Fe(9d)=Fe2 | 2.26 | 1.97 | 1.54 |
| Fe(18f)=Fe3 | 2.17 | 2.15 | - |
| Fe(18h)=Fe4 | 2.31 | 2.18 | 2.10 |
| **Total [$\mu_B$/f.u.]** | 23.7 | 14.8 | 6.3 |

**Parameters of nearest-neighbor exchange interactions and $T_c$**

The results for the exchange interaction parameters $I_{ab}(\mathbf{r}_a-\mathbf{r}_b)$ between different pairs of nearest-neighbor iron ions (in units of K) are presented in Table 3. Here $I_{33}(1)$ stands for the value of the exchange parameter between ions Fe3 and Fe3 at the nearest distance between these ions (index 1 in the brackets) for an edge of hexagon, then $I_{33}(2)$ denotes the same type interaction at the second order distance in another crystallographic direction (another edge of unequalateral hexagon).

Table 3. Parameters of exchange in the rhombohedral structure of $Gd_2Fe_{1-x}Ga_x$ for the first coordination sphere.

| Types of Exchanges | Exchange [K] for $Gd_2Fe_{17}$ | Exchange [K] for $Gd_2Fe_{14}Ga_3$ | Exchange [K] for $Gd_2Fe_{17}Ga_6$ |
|---|---|---|---|
| $I_{11}(1)$ | 287.5 | 433.6 | 436.1 |
| $I_{44}(1)$ | 182.2 | 187.1 | 259.6 |
| $I_{34}(1)$ | 125.9 | 184.8 | - |
| $I_{33}(2)$ | 121.9 | - | - |
| $I_{24}(1)$ | 121.0 | 246.8 | 237.8 |
| $I_{34}(2)$ | 105.7 | 11.5 | - |
| $I_{14}(1)$ | 88.8 | 78.2 | 151.7 |
| $I_{23}(1)$ | 87.1 | 188.2 | - |
| $I_{12}(1)$ | 83.6 | 120.4 | 165.8 |
| $I_{13}(1)$ | 74.1 | 132.3 | - |
| $I_{33}(1)$ | -36.5 | - | - |

From this data it is clear that doping of Ga gives rise to exchange parameters between Fe atoms (in spite of increasing distances (see Table 4)), what can be explained with the additional amount of superexchange interaction between magnetic ions through nonmagnetic ions of Ga. At the same time another tendency is obvious – the number of exchange interactions between Fe atoms goes down with doping as the number of Fe atoms decrees itself. So these too tendencies compete with each other. And it is reasonable to use the Curie point $T_c$ to estimate collective effect of these changes.

Table 4. Distances between atoms in the first coordination sphere for rhombohedral structure of $Gd_2Fe_{1-x}Ga_x$

| Types of Exchanges | Distance [A] for $Gd_2Fe_{17}$ | Distance [A] for $Gd_2Fe_{14}Ga_3$ | Distance [A] for $Gd_2Fe_{17}Ga_6$ |
|---|---|---|---|
| $I_{11}(1)$ | 2.397 | 2.421 | 2.439 |
| $I_{44}(1)$ | 2.502 | 2.525 | 2.548 |
| $I_{34}(1)$ | 2.563 | 2.589 | - |
| $I_{33}(2)$ | 3.581 | - | - |
| $I_{24}(1)$ | 2.460 | 2.483 | 2.506 |
| $I_{34}(2)$ | 2.625 | 2.651 | - |
| $I_{14}(1)$ | 2.652 | 2.677 | 2.701 |
| $I_{23}(1)$ | 2.435 | 2.459 | - |
| $I_{12}(1)$ | 2.615 | 2.639 | 2.663 |
| $I_{13}(1)$ | 2.753 | 2.779 | - |
| $I_{33}(1)$ | 2.479 | - | - |

We solve the set of equations [10], obtained in the frame of Weiss mean-field theory:

$$k_B T_c \sigma_a = \frac{S_a^2}{3} \sum_b \sum_{|\Delta_{ab}|} I_{ab}(|\Delta_{ab}|) z_{ab}(|\Delta_{ab}|) \sigma_b , \qquad (1)$$

where $\sigma_b = \langle S_b^Z(r_a+\Delta_{ab}) \rangle$ denotes the thermodynamic average of z-projection of classical spin, and a,b=1,2,3,4 for all Fe ions, $z_{ab}$ – number of nearest neighbors.

The estimations of the Curie point are $T_c(Gd_2Fe_{17})$=429K, $T_c(Gd_2Fe_{14}Ga_3)$=521K and $T_c(Gd_2Fe_{11}Ga_6)$=381K, when the experimental results for the same concentrations are [4]: $T_c(Gd_2Fe_{17})$=465K, $T_c(Gd_2Fe_{14}Ga_3)$=610K and $T_c(Gd_2Fe_{11}Ga_6)$=410K (see in Table 1). On Fig.4 one can see experimental points [4] denoted by black circles. From that curve it is seen that $T_c$ grows with concentration up to doping x=3, and from x=3 to x=6 it goes down. Obtained by us $T_c$ points are denoted by red rhombs on Fig4. They represent the same kind of doping dependence. It can be explained by the fact that values of exchange parameters $I_{ab}$ rise from x=0 to x=3 because of additional amount of superexchange interaction through the nonmagnetic atoms of Ga. From x=3 to x=6, in spite of the fact that values of exchange interaction still grow, the number of exchange pairs decreases that summarize in the decrease of $T_c$.

Though we obtain the tendency of Curie temperature behavior by our calculations, but to achieve precise fit we should fulfill several conditions: first, to add the exchanges between ions of remote coordination spheres; second, to introduce temperature dependence of exchange parameters, since in the present calculations we use exchange parameters obtained for the ground state of $Gd_2Fe_{1-x}Ga_x$.

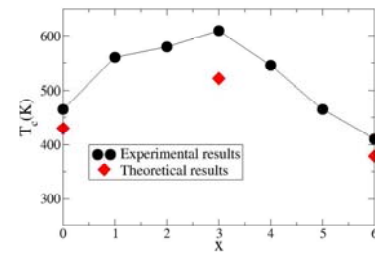

Fig.4. Concentration dependence of Curie temperature

**Summary**

In the framework of LSDA method electronic structures, magnetic moments and exchange interactions were calculated in the first coordination sphere of subsystem of Fe - ions of the compounds: $Gd_2Fe_{17}$, $Gd_2Fe_{14}Ga_3$ and $Gd_2Fe_{11}Ga_6$.

It was shown that doping with Ga atoms gives a significant growth of exchange ferromagnetic parameters between nearest Fe ions, which can be connected with additional superexchange interactions appearing between Fe ions through Ga atoms. At the same time the number of exchange-connected pairs of Fe atoms goes down with doping together with exchange contribution to the free energy. Thus the interference of that two opposite tendencies leads to nonmonotonic $T_c$ behavior with increase of Ga amount with maximum near x=3.

Estimation of $T_c$ concentration dependence in the framework of molecular field approximation with the calculated parameters of exchange interaction for nearest neighbors Fe ions gives the following values: $T_c(Gd_2Fe_{17})$=429K, $T_c(Gd_2Fe_{14}Ga_3)$=521K and $T_c(Gd_2Fe_{11}Ga_6)$=381K. Though those values are lower then experimental ones, they nicely reproduce an observed experimental nonmonotonic concentration $T_c$ dependence [4].

This work is partly supported by RFBR grant 08-02-00021, programs of fundamental research of the Russian Academy of Sciences (RAS) ``Quantum physics of condensed matter'' (09-П-2-1009) and of the Physics Division of RAS ``Strongly correlated electrons in solid states'' (09-T-2-1011). IN thanks Grant of President of Russia MK-614.2009.2, interdisciplinary UB-SB RAS project.